\documentstyle[twoside,12pt,epsf]{article}
\normalsize
\newcommand{\bdi}{\begin{displaymath}}
\newcommand{\edi}{\end{displaymath}}
\newcommand{\bfi}{\begin{figure}}
\newcommand{\efi}{\end{figure}}

\newcommand{\beq}{\begin{equation}}
\newcommand{\eeq}{\end{equation}}
\newcommand{\beqa}{\begin{eqnarray}}
\newcommand{\eeqa}{\end{eqnarray}}

\newcommand{\ra}{\rightarrow}

\def\longbar#1{\setbox1=\hbox{$#1$}
\setbox2=\vbox{\hrule width 0.8\wd1}
\raise0.5\ht1\hbox{${\lower\dp1\box2}\atop\box1$}}  

\begin{document}

\begin{titlepage}

\begin{flushright}
\today
\end{flushright}

\vspace{1cm}
\begin{center}
{\Large \bf Degeneracy of zero modes of the Dirac operator in three
dimensions}\\[1cm]
C. Adam* \\
School of Mathematics, Trinity College, Dublin 2 \\

\medskip

\medskip

B. Muratori**,\, C. Nash*** \\
Department of Mathematical Physics, National University of Ireland, Maynooth
\vfill
{\bf Abstract} \\
\end{center}
One of the key properties of Dirac operators is the possibility of 
a degeneracy of zero modes. For the Abelian Dirac operator in three
dimensions the question whether such multiple zero modes may exist
has remained unanswered until now. Here we prove that the feature of
zero mode degeneracy indeed occurs for the Abelian Dirac operator in
three dimensions, by explicitly constructing a class of Dirac
operators together with their multiple zero modes. Further, we discuss
some implications of our results, especially a possible relation to the
topological feature of Hopf maps.  
\vfill

$^*)${\footnotesize  
email address: adam@maths.tcd.ie, adam@pap.univie.ac.at} 

$^{**})${\footnotesize
email address: bmurator@fermi1.thphys.may.ie} 

$^{***})${\footnotesize
email address: cnash@stokes2.thphys.may.ie} 
\end{titlepage}

\section{Introduction}

Fermionic zero modes of
the Dirac operator $D_A = \gamma^\mu (\partial_\mu -iA_\mu )$ are of 
importance in many places in quantum field theory and mathematical
physics \cite{AS1,JR1,JR2}. They are 
the ingredients for the computation of the index of the Dirac operator
and play a key r\^ole in understanding anomalies. In Abelian gauge 
theories, which is what we are concerned with here, they affect 
crucially the behaviour of the Fermion determinant $\det(D_A)$ 
in quantum electrodynamics. The nature  of the QED 
functional integral depends strongly on the degeneracy of the 
bound zero modes.

In three dimensions -- which is the case which we want to study here --
the first examples of such zero energy Fermion bound states have been
obtained only in 1986 \cite{LoYa1}, and some further results have been found 
recently \cite{zero}. In both articles no degeneracy of these zero modes
has been observed, because, by their very methods, the authors of 
\cite{LoYa1} and of \cite{zero} could only construct one zero mode
per gauge field. Hence, the question of a possible degeneracy of 
zero energy bound states of the Abelian Dirac operator in three dimensions 
has remained completely unanswered up to now.

  It should be emphasized here that the problem of the existence and
degeneracy of zero modes of the
Abelian Dirac operator in three dimensions, 
in addition to being interesting in its own right, has
some deep physical implications. The authors of \cite{LoYa1} were mainly
interested in these zero modes because in an accompanying paper
\cite{FLL} it was proven that one-electron atoms with sufficiently high
nuclear charge in an external magnetic field are unstable if such zero modes 
of the Dirac operator exist. 

Further, there is an intimate connection between the existence and
number of zero modes of the Dirac operator for strong magnetic fields on
the one hand, and the nonperturbative 
behaviour of the three dimensional Fermionic
determinant (for massive Fermions) in strong external magnetic fields on the
other hand. The behaviour of this determinant, in turn, is related to
the paramagnetism of charged Fermions, see \cite{Fry1,Fry2}.
So, a thorough understanding of the zero modes of the Dirac operator
is of utmost importance for the understanding of some deep physical
problems as well.

In addition, 
it is speculated in \cite{Fry2} that the existence 
and degeneracy of zero modes  for $QED_3$ may  have a topological origin 
as it does in $QED_2$ \cite{Jac1}--\cite{Adam} --- cf. 
\cite{Fry2} for details and an account 
of the situation for $QED_{2,3,4}$.

It is the purpose of this letter to address the question of a possible
degeneracy of zero energy bound states of the Abelian Dirac operator in
three dimensions. We shall prove that the phenomenon of a degeneracy of
zero modes does indeed occur, by constructing a special class of gauge
fields that lead to an arbitrary number of square-integrable zero modes
of the corresponding Dirac operators.
In addition, we shall discuss how our results are related to some
rigorous bounds on the number of zero modes, and we shall indicate a possible
relation of our results to some underlying topological properties of the
special class of gauge fields that we provide.

\section{Construction of the zero modes}

We are interested in solutions of the three-dimensional, Abelian Dirac equation
(the Pauli equation)
\beq
-i\sigma_i \partial_i \Psi (x) 
=A_i (x) \sigma_i \Psi (x).
\eeq
Here $\vec x =(x_1 ,x_2 ,x_3)^{\rm T}$, $i,j,k = 1\ldots 3$, $\Psi$ is a
two-component, square-integrable spinor on ${\rm\bf R}^3$, $\sigma_i$
are the Pauli matrices and $A_i$ is an Abelian gauge field. The
authors of \cite{LoYa1} observed that a solution to this equation could
be obtained from a solution to the simpler equation
\beq
-i\vec\sigma \vec\partial \Psi =h\Psi
\eeq
for some scalar function $h(x)$. In this case the corresponding gauge field
that obeys the Dirac equation (1) together with the spinor (2) is given
by
\beq
A_i =h \frac{\Psi^\dagger \sigma_i \Psi}{\Psi^\dagger \Psi} .
\eeq
In addition, they gave the following explicit example
\beq
\Psi =(1+r^2)^{-\frac{3}{2}}({\bf 1} +i\vec x \vec \sigma )\Phi_0
\eeq
where $\Phi_0$ is the constant unit spinor $\Phi_0 =(1,0)^{\rm T}$. The
spinor (4) obeys
\beq
-i \vec\sigma \vec\partial  \Psi = \frac{3}{1+r^2}\Psi
\eeq
and is, therefore, a zero mode for the gauge field
\beq
\vec A = \frac{3}{1+r^2} \frac{\Psi^\dagger \vec\sigma \Psi}{\Psi^\dagger \Psi}
=\frac{3}{(1+r^2)^2} 
\left( \begin{array}{c} 2x_1 x_3 -2x_2  \\ 2x_2 x_3 +2x_1 \\
1-x_1^2 -x^2_2 +x_3^2 \end{array} \right) .
\eeq
The magnetic field $\vec B =\vec\partial\times \vec A$ for the gauge field (6)
reads
\beq
\vec B 
=\frac{12}{(1+r^2)^3} 
\left( \begin{array}{c} 2x_1 x_3 -2x_2  \\ 2x_2 x_3 +2x_1 \\
1-x_1^2 -x^2_2 +x_3^2 \end{array} \right) .
\eeq
Now assume that a function $\chi$ exists such that
\beq
(-i\sigma_j \partial_j \chi )({\bf 1} +i\vec x \vec \sigma )\Phi_0 =0
\eeq
then $\chi^n \Psi$, $n\in {\rm \bf Z}$ (where $\Psi$ is the zero mode (4)),
are additional formal zero modes for the same gauge field (6).
Condition (8) implies
\beq
{\rm det} (-i\vec\sigma\vec\partial\chi ) =\sum_{i=1}^3 \chi_{,i}\chi_{,i}
=0,
\eeq
therefore, $\chi$ necessarily must be complex. Indeed, such a function
$\chi$ fulfilling (8) exists,
\beq
\chi =:Se^{i\sigma} = \frac{2(x_1 +ix_2)}{2x_3 -i(1-r^2)}
\eeq
\beq
S^2 = \frac{4(r^2 -x_3^2)}{4x_3^2 +(1-r^2)^2}\, ,\quad
\sigma ={\rm atan}\, \frac{x_2}{x_1} + {\rm atan}\, \frac{1-r^2}{2x_3} .
\eeq
as may be checked easily. Here we expressed $\chi$ by its modulus $S$ and
phase $\sigma$ for later convenience. For the formal zero modes $\chi^n
\Psi$ we observe the following two points. Firstly, $n$ has to be 
integer, because only integer powers of $\chi$ lead to a single-valued
spinor $\chi^n \Psi$. Secondly, $\chi^n \Psi$ is singular for all
$n\in {\rm\bf Z}\setminus \{ 0\}$, because $\chi$ is singular along the circle
$\{ \vec x \in {\rm\bf R}^3 \, \backslash \, x_3 = 0 \, 
, x_1^2 + x_2^2 =1\}$ and
zero along the $x_3$ axis. Therefore, the formal zero modes $\chi^n \Psi$,
with $\Psi$ given in (4), are not acceptable. However, we shall find some
zero modes, different from (4), where multiplication with $\chi^n$ will
lead to acceptable new zero modes for some $n\neq 0$. For this purpose, let us
first review some more results of \cite{LoYa1}. 
The authors of \cite{LoYa1} observed that, in addition to their simplest 
solution (4), they could find similar solutions to eq. (2) with higher 
angular momentum. Using instead of the constant spinor $\Phi_0 =(1,0)^{\rm T}$
the spinor
\beq
\Phi_{l,m}= \left( \begin{array}{c} \sqrt{l+m+1/2}\, Y_{l,m-1/2} \\
 -\sqrt{l-m+1/2}\, Y_{l,m+1/2} \end{array} \right)
\eeq
(where $m\in [-l-1/2\, ,\, l+1/2]$ and $Y$ are spherical harmonics), 
they found the solutions
\beq
\Psi_{l,m}=r^l (1+r^2)^{-l-\frac{3}{2}}({\rm\bf 1}+
i\vec x \vec \sigma )\Phi_{l,m}
\eeq
\beq
\vec A_{l,m}=(2l+3)(1+r^2)^{-1}\frac{\Psi^\dagger_{l,m}\vec\sigma
\Psi_{l,m}}{\Psi^\dagger_{l,m} \Psi_{l,m}} .
\eeq
Specifically, for maximal magnetic quantum number $m=l+1/2$, these solutions
read
\beq
\Psi_l :=\Psi_{l,l+1/2} = \frac{Y_{l,l} r^l}{(1+r^2)^{l+3/2}}
({\rm\bf 1}+ i\vec x \vec \sigma )\Phi_0
\eeq
\beq
\vec A^{(l)} 
=\frac{3+2l}{(1+r^2)^2} 
\left( \begin{array}{c} 2x_1 x_3 -2x_2  \\ 2x_2 x_3 +2x_1 \\
1-x_1^2 -x^2_2 +x_3^2 \end{array} \right) 
\eeq
\beq
\vec B^{(l)}  
=\frac{4(3+2l)}{(1+r^2)^3} 
\left( \begin{array}{c} 2x_1 x_3 -2x_2  \\ 2x_2 x_3 +2x_1 \\
1-x_1^2 -x^2_2 +x_3^2 \end{array} \right) 
\eeq
(where we have omitted an irrelevant constant factor in (15)).
Hence, $\Psi_l$ is proportional to the simplest zero mode (4) and is, 
therefore, still an eigenvector of the matrix $-i\sigma_j \partial_j \chi$
with eigenvalue zero. Further, 
the zero mode $\Psi_l$ may be rewritten as (again, we ignore irrelevant 
constant factors)
\beq
\Psi_l = e^{il\varphi}\frac{S^l}{(1+S^2)^{l/2}}
(1+r^2)^{-3/2}
({\rm\bf 1}+ i\vec x \vec \sigma )\Phi_0
\eeq
where we introduced polar coordinates $(x_1 ,x_2 ,x_3) \ra (r,\theta ,
\varphi)$,
$S$ is the modulus (11), and 
\beq
Y_{l,l} = e^{il\varphi}\sin^l \theta =e^{il\varphi}\frac{(r^2 -
x_3^2)^{l/2}}{r^l}
= e^{il\varphi}\frac{(1+r^2)^l}{r^l}\frac{S^l}{(1+S^2)^{l/2}} .
\eeq
From expression (18) for $\Psi_l$ it follows easily that the spinors
\beq
\Psi_{n,l}=\chi^{-n}\Psi_l = e^{i(l\varphi -n\sigma)}
\frac{S^{l-n}}{(1+S^2)^{l/2}} 
\Psi \, , \quad n=0,\ldots l
\eeq
(where $\Psi$ is the spinor (4) and $\sigma$ is the phase (11)),
are non-singular, square-integrable zero modes for the same gauge field
$\vec A^{(l)}$ and, therefore, the Dirac operator with gauge field
$\vec A^{(l)}$ given by (16) has $l+1$ square-integrable zero modes.

\section{Discussion}

We explicitly constructed the gauge fields $\vec A^{(l)}$, (16), and showed 
that the corresponding Dirac operator has $l+1$ non-singular, square-integrable
zero energy bound states (20). 
Hence we proved by explicit construction that the
phenomenon of zero mode degeneracy does occur for the Abelian Dirac operator
in three dimensions, which has been unknown until now.

Before closing, we want to further comment on some points. Firstly, the
magnetic fields (17) for higher $l$ are just multiples of the simplest magnetic
field (7). Therefore, the number of zero modes $N_l$ for strong magnetic fields
(i.e. large $l$) behaves like
\beq
N_l = l+1 \sim c\int d^3 x |\vec B^{(l)}|
\eeq
(it holds that $\lim_{|\vec x| \to \infty}|\vec B^{(l)}| \sim r^{-4}$, 
therefore the integral in (21) exists),
i.e., $N_l$ grows linearly with the strength of the magnetic field (here $c$
is some constant). This is well within the rigorous upper bound on the
possible growth of the number of zero modes
\beq
N\sim c' \int d^3 x |\vec B|^{3/2}
\eeq
that was first stated in \cite{LoYa1} and later derived in \cite{Fry2} 
(here $c'$ is a constant).

Secondly, there is in fact a relation between our magnetic fields (17) and
the topological feature of Hopf maps. Hopf maps are maps $S^3 \ra S^2$,
and they fall into distinct homotopy classes that are labelled by the
integers (the Hopf index). Hopf maps may be represented by complex
functions $\chi : {\rm\bf R}^3 \ra {\rm\bf C}$ provided that $\chi (
|\vec x|=\infty) = \chi_0 ={\rm const}$. Here the coordinates in ${\rm\bf R}^3$
and ${\rm\bf C}$ are interpreted as stereographic coordinates of the $S^3$
and $S^2$, respectively. Further, a magnetic field $\vec B$ (the Hopf
curvature) is related to each Hopf map, where $\vec B$ is tangent to the
closed curves $\chi ={\rm const}$, see \cite{Ran1}--\cite{hoin} for details.

The observation which we want to make here is that the function $\chi$ in 
(10) is just the simplest standard Hopf map with Hopf index 1, and the 
magnetic field (7) is related to the corresponding Hopf curvature of 
$\chi$. More precisely, after the addition of a fixed prescribed background
magnetic field, the magnetic field (7) is precisely equal to the Hopf 
curvature of the simplest Hopf map $\chi$ (see \cite{hoin}). 
Further, as the higher $\vec B^{(l)}$ are related to the simplest 
$\vec B$ by integer coefficients, (17), they are related to 
the Hopf curvatures of higher Hopf maps in the same way (i.e., they are
equal to higher Hopf curvatures after the addition of the same fixed
background magnetic field). The corresponding Hopf index $H_l$ may be 
expressed by the number of zero modes $N_l$ as
\beq
H_l = ((N_l +1)/2)^2
\eeq
(when the conventions of \cite{hoin} for the Hopf curvature are used). 
Eq. (23) will be proved and discussed in more detail elsewhere, \cite{AMN1}.
As $l$ is related to the angular momentum, (13), this indicates an
interesting connection between angular momentum, the number of zero modes 
and the Hopf index.

The above observation leads, of course, to the question whether the
topological feature of Hopf maps is related to the existence and degeneracy
of zero modes in the general case. 
This question will be investigated further in
future publications. Here we just want to mention that there exist more
zero modes that are proportional the simplest zero mode (4) up to a
scalar function, see e.g. \cite{hoin}.
For these zero modes it remains true that additional 
formal zero modes of the same
Dirac operator may be constructed by multiplication with the complex
function $\chi$, as we did above. In addition, the number of
square-integrable zero modes remains related to the Hopf index of the
corresponding magnetic field (when interpreted as a Hopf curvature) as
in (23), \cite{AMN1}.

Anyhow, we think that our results will be relevant for some future
developments
in mathematical physics, as well as for the understanding of 
non-perturbative aspects of quantum
electrodynamics, especially in three dimensions.   

\section{Acknowledgments}
The authors thank M. Fry for helpful discussions. In addition,
CA gratefully acknowledges useful conversations with R. Jackiw.
CA is supported by a Forbairt Basic Research Grant.
BM gratefully acknowledges financial support from the Training and 
Mobility of Researchers scheme (TMR no. ERBFMBICT983476).

\end{document}